# Robustifying deep networks for image segmentation

Zheng Liu[1], Jinnian Zhang[2], Varun Jog[2], Po-Ling Loh[1,2], Alan B McMillan[3,4*]

[1]Department of Statistics, [2]Department of Electrical and Computer Engineering, [3]Department of Medical Physics, [4]Department of Radiology, University of Wisconsin, Madison ([*]abmcmillan@wisc.edu)

**Abstract**

Purpose: The purpose of this study is to investigate the robustness of a commonly-used convolutional neural network for image segmentation with respect to visually-subtle adversarial perturbations, and suggest new methods to make these networks more robust to such perturbations. Materials and Methods: In this retrospective study, the accuracy of brain tumor segmentation was studied in subjects with low- and high-grade gliomas. A three-dimensional UNet model was implemented to segment four different MR series (T1-weighted, post-contrast T1-weighted, T2- weighted, and T2-weighted FLAIR) into four pixelwise labels (Gd-enhancing tumor, peritumoral edema, necrotic and non-enhancing tumor, and background). We developed attack strategies based on the Fast Gradient Sign Method (FGSM), iterative FGSM (i-FGSM), and targeted iterative FGSM (ti-FGSM) to produce effective attacks. Additionally, we explored the effectiveness of distillation and adversarial training via data augmentation to counteract adversarial attacks. Robustness was measured by comparing the Dice coefficient for each attack method using Wilcoxon signed-rank tests. Results: Attacks based on FGSM, i-FGSM, and ti-FGSM were effective in significantly reducing the quality of image segmentation with reductions in Dice coefficient by up to 65%. For attack defenses, distillation performed significantly better than adversarial training approaches. However, all defense approaches performed worse compared to unperturbed test images. Conclusion: Segmentation networks can be adversely affected by targeted attacks that introduce visually minor (and potentially undetectable) modifications to existing images. With an increasing interest in applying deep learning techniques to medical imaging data, it is important to quantify the ramifications of adversarial inputs (either intentional or unintentional).

**Introduction**

Machine learning algorithms have become increasingly popular in medical imaging [1-3], where highly functional algorithms have been trained to recognize patterns in image data sets and perform clinically relevant tasks such as tumor segmentation and disease diagnosis. In particular, approaches based on deep learning have recently drawn widespread attention[4]. However, an oft-repeated criticism of deep learning is that it uses a "black-box" approach, giving rise to decision-making processes that are uninterpretable even by domain experts and deep learning researchers [5][6]. Furthermore, it has been demonstrated that neural networks can be tricked into misclassifying images when they are perturbed by negligible amounts of specific types of noise [7].

The vast majority of existing deep learning theory on robustness focuses on classification tasks, where the goal of an adversarial attack is to generate a small perturbation to an image that causes a highly-accurate neural network to misclassify the perturbed image. In contrast, image segmentation tasks are more complex. Although image segmentation may be viewed as a classification procedure operating on the individual pixels of an image, adversarial attacks typically consist of applying a global perturbation to the entire image that simultaneously changes the classes of all pixels in an adversarial manner. Additional work is necessary to study which global perturbations might lead to specific types of segmentation errors, particularly for convolutional neural networks (CNNs), which are nearly ubiquitous in deep learning applications for medical image processing.

In this paper, we investigate the vulnerability of deep learning algorithms in the context of image segmentation, and propose methods that may be adopted during training to make deep learning algorithms more robust. Our objective is to study the hypothesis that adversarial attack strategies developed for neural network classifiers using visually subtle perturbations to input images [7][8] can be adapted to the task of medical image segmentation. Our second contribution is to investigate methods for defending against such adversarial attacks. It has been shown in recent studies that adversarial training [8] and defensive distillation [9] can increase the robustness of neural networks when applied to classification tasks for standard computer vision datasets such as MNIST and ImageNet [16]. In this paper, we will show that these methods are also effective for medical imaging segmentation. Both of our defense strategies, based on adversarial training and defensive distillation, show significantly improved robustness with respect to adversarial attacks.

**Materials and Methods**

Data from the The Cancer Imaging Archive (TCIA) and the Medical Image Computing & Computer Assisted Intervention (MICCAI) Brain Tumor Segmentation (BraTS) 2017 challenge [17-20] were used for this IRB-exempt study. These publicly-available, retrospective datasets from multi-institutional studies consist of magnetic resonance (MR) images of the brain from 283 subjects with either low-grade glioma or glioblastoma multiforme. Each dataset includes four different MR series: (a) T1-weighted [T1], (b) post-contrast T1-weighted [T1Gd], (c) T2-weighted [T2], and (d) T2-weighted Fluid Attenuated Inversion Recovery [FLAIR]. Segmentation volumes, manually segmented by expert neuroradiologists, are provided with the following pixelwise labels: (i) Gd-enhancing tumor (ET-label 4), (ii) peritumoral edema (ED-label 2), (iii) necrotic and non-enhancing tumor (NCR/NET-label 1), and (iv) background (label 0). We reserved 20% of the data for testing. We used a popular CNN architecture, the 3D-Unet model [10], which has demonstrated good performance for segmentation tasks on this dataset and is described in detail in the Supplemental Materials.

*Attacks*

We focused on first-order attacks, which construct perturbations based on the gradient of a loss function evaluated on input images in a trained network. We also studied targeted attacks, which encourage the result of a perturbation to fall into a specific category. For example, given an input tumor image, the goal might be to construct an adversary that results in an output that moves the tumor label to a certain (incorrect) position. We developed attack strategies based on Fast Gradient Sign Method (FGSM) [1], iterative FGSM (i-FGSM) [8], and targeted iterative FGSM (ti-FGSM) [8].

Mathematical details for our procedures may be found in the Supplemental Materials. A key idea in our approach was to replace the cross-entropy loss used for usual classification tasks with the Dice coefficient loss in the FGSM algorithm. The Dice coefficient loss is equal to $(1 -$ Dice coefficient$)$ of the model segmented image and the true output, where the Dice coefficient is a metric which assesses the spatial overlap of two image segmentations [21]. For $\varepsilon$, we chose 5% of the maximum pixel magnitude of the input image. We chose $\alpha$ and the number of steps $N$ such that $\alpha N = 5\%$ of the maximum pixel magnitude of the input image.

*Defenses*

For defenses, we first explored a method based on distillation [9]. The key idea is to retrain a neural network on a dataset using vectors of soft labels that are obtained from an initial training stage of the neural network. The classification function of the "distilled" neural network, which predicts soft label vectors from inputs, is a continuous function that is smoother over the domain of input variables than the original network, thus is less sensitive to small input variations. A temperature parameter T, which controls the desired level of smoothness of the distilled network, is introduced to the activation function of the output layer during training. All mathematical details of our adaptation of defensive distillation for image segmentation may be found in the Supplemental Materials.

Recent work [25-27] has suggested that data augmentation, which introduces artificially generated images to the training set by adding small amounts of random noise to training images (e.g., from uniform or Gaussian distributions on pixel magnitudes [25, 26], or random rotations to the overall image [27]), also produces more robust networks. However, previous literature also suggests that data augmentation may have a limited benefit to adversarial robustness. We compared our distillation strategy to the performance of the simpler data augmentation technique. Additional training details may be found in the Supplemental Materials.

*Measuring Robustness*

To study the effects of adversarial attacks, we used fixed values of $\varepsilon$. For FGSM, we chose $\varepsilon$ to be 0.05, which corresponds to 5% of the maximum intensity of the image. For i-FGSM, we chose $\varepsilon = 0.005$ and the number of iterates to be 10. For ti-FGSM, the target was all labels in the image equal to 1 (i.e., necrotic and non-enhancing tumor). To study the effects of network defenses, we used a range of $\varepsilon$ values from 0 to 0.010, in increments of 0.001.

We evaluated the robustness of the attacked and defended networks by quantifying the effect of adversarial perturbations. The overall robustness of a classifier was obtained by comparing the average Dice coefficient of the segmented, adversarially perturbed test images with the average Dice coefficient of segmented, non-perturbed images. Wilcoxon signed-rank tests were used to determine whether the proposed perturbation strategies for inputs resulted in significantly different Dice coefficients of segmented outputs. Similarly, the Peak Signal-to-Noise Ratio (PSNR), Structural Similarity Index (SSIM), and root mean squared error (RMSE) of the perturbed input images were compared to the ground truth images for each type of attack.

**Results**

Analysis was performed on data from 283 datasets. Demographic data is not available for this dataset; however for 163 of the subjects age (60.3±12.1 years) and overall survival (423±350 days) was available.

*Adversarial attacks*

Example adversarial attacks are shown in Fig. 1, where we see that all three adversaries successfully inject errors into the segmented images, with minimal visual disturbance to the input images. This verifies that small adversarial perturbations to the input image can indeed have a substantial impact on the resulting segmentation.

The average Dice coefficients (mean ± standard deviation) of the predicted output with respect to the ground-truth masks and the PSNR, SSIM, and RMSE of the input images are shown in Table 1. A Wilcoxon signed-rank test was used to compare the Dice coefficient to the ground truth data for each attack type ($p \leq 0.05$). A Bonferroni correction was applied to correct for multiple comparisons. The attacks are highly successful, since all variants of FGSM result in a significantly lower Dice coefficient. Compared to the "No attack" condition, the attacks reduced the Dice coefficient by 30.5%, 58.3%, and 43.8% in the tumor core; 44.6%, 65.6%, and 45.4% in the enhancing tumor; and 26.7%, 47.5%, and 35.0% in the whole tumor for the FGSM, i-FGSM, and ti-FGSM, respectively. Despite visually subtle changes, image quality metrics PSNR and SSIM suggest measurable differences between input images, while RMSE differences are low.

In Fig. 2, we show plots of the average Dice coefficient vs. number of iterations in i-FGSM and ti-FGSM. As expected, with an increasing number of iterations, we see a steadily decreasing Dice coefficient—indicating that with more steps, the adversaries become stronger, causing the segmentation output to worsen. The effects of i-FGSM iterations on image input quality are shown in the Supplemental Materials.

*Defense via distillation*

The performance of distilled neural networks for different temperatures is shown in Fig. 3. In each plot, the robustness of the neural network clearly increases with T. For T=5000, the gains are 0.14, 0.27, and 0.22, respectively, compared to the network without distilled training (T=1) at the worst attack case. This indicates that distillation is indeed effective in defending against the proposed adversarial attacks.

It is also notable that the improvement appears to saturate when the temperature exceeds a certain threshold. For example, gains in robustness for temperatures over 100 in Fig. 3(a) are negligible. The phenomenon is not observed in Figs. 3(b) and (c), because the threshold for the temperature is higher than in (a).

Moreover, we observe that increasing the temperature can make neural networks more robust, while maintaining a testing accuracy that is comparable to the original model, corroborating previous findings [9]. Defensive distillation also has the potential to improve testing accuracy [9]: This phenomenon is more obvious in (c), in which all distilled networks outperform the original model when $\varepsilon$ is equal to 0. The main drawback of using a larger temperature is slower convergence while training, leading to a higher computational workload. This may impose practical constraints on the magnitude of T that can be used while training.

*Adversarial training*

Fig. 4 shows the Dice coefficients of different models by using adversarial training with different values of $\varepsilon$. For all categories, adversarial training is seen to enhance the robustness of neural networks. For comparison, we also plot the line (marked with stars) of data augmentation with random perturbations of radius of 0.01 in infinity norm [25-27]. This leads to better robustness than the original neural network; however, the starred line lies below all other lines, indicating that more sophisticated defenses will make the trained networks more robust.

Similar to defensive distillation, different values of $\varepsilon$ used in adversarial training only have moderate effects on the test accuracy, which may be seen by comparing the curves in each category when $\varepsilon$ is 0. However, when we evaluate the performance of each model across all categories, the increase of $\varepsilon$ in the training process does not ensure improved robustness. Moreover, the training process may diverge for large values of $\varepsilon$, making the choice of $\varepsilon$ crucial.

Fig. 5 shows an example of adversarial images of different models and the corresponding predicted labels. The leftmost column contains the original image and its true label. Note that all models perform well on the unperturbed images, since the Dice coefficient for label = 4 (Enhancing Tumor) is around 0.70. Next, we apply FGSM with $\varepsilon = 0.03$ to generate adversarial images, which are shown in the middle row. We can see that the perturbations are nearly imperceptible to the human eye. However, the Dice coefficients in the 5$^{th}$ and 6$^{th}$ columns (model with no defense, and distilled model with T=20) dropped down significantly in the 3$^{rd}$ row, while the others remain almost the same. A Wilcoxon signed-rank test was used to compare the Dice coefficient to the ground truth data for each attack type (*p*≤0.05), and a Bonferroni correction was applied to correct for multiple comparisons. A summary of the performance of these models on the testing dataset can be found in Table 2. Notably, although these defensive models achieve better performance on adversarial examples, they still perform worse than the models applied to unperturbed images.

*Distillation vs. adversarial training*

Based on the results in Table 2 and compared to the "No defense" condition with $\varepsilon$ = 0 (No attack), distillation with T=5000 performed the best for $\varepsilon$ > 0, and distillation with T=20 performed the worst for each tumor segmentation region. However, it is not necessarily true that defensive distillation will *always* outperform adversarial training in terms of a one-step attack. With a more careful choice of $\varepsilon$, the performance of adversarial training may exceed that of distillation; however, it may be more difficult to find the optimal choice of $\varepsilon$, compared to tuning the temperature to obtain better performance.

**Discussion**

We have demonstrated the vulnerability of deep learning algorithms for image segmentation tasks to adversarial perturbations. Adversarial attacks create imperceptible visual differences to the input data, yet have profound effects on the segmented output. Furthermore, we have developed methods for easily constructing adversarial perturbations using generalizations of FGSM, and have similarly studied defense mechanisms based on distillation and adversarial training. We have illustrated the effectiveness of our methods on medical images in the BraTS dataset.

In the current work, we have mainly focused on one-step adversarial attacks. Integrating more sophisticated adversaries during training is likely to make the networks more robust, and constitutes part of our future work. Furthermore, recent work [8] shows that adversarial training may result in label "leak" if the original task is difficult, such as classification tasks on the ImageNet dataset. Label leak occurs when a model is trained using adversarial attacks generated by FGSM and again evaluated using images with FGSM perturbations, producing higher accuracy on adversarial examples than on clean images. A potential explanation is that the gradient added to the original image in adversarial training contains extra information from the label, making classification of adversarial examples easier if a neural network uses that information. We plan to investigate whether label leak also occurs for segmentation and classification tasks in medical imaging. Lastly, we plan to evaluate the effectiveness of different defense techniques beyond standard white-box attacks on the trained model. For instance, we are interested in examining whether a defense strategy is effective against black-box adversarial examples or transferred adversarial attacks [23][24].

This work is not without limitations. First, we have focused on a certain network structure, a 3D-Unet model. It would be interesting to see if other network structures also lead to similar trends with respect to adversarial attacks and defenses—perhaps specific network structures could be designed to increase robustness to certain types of attacks. Second, we have mainly studied adversarial attacks based on FGSM, since they generate adversarial perturbations in a fast, simple way, However, one could similarly adapt other attack methods such as Deepfool [12], JSMA [13], and DAG [15] from classification to segmentation tasks. These methods could lead to more effective attacks, particularly for targeted attack strategies, where our results show that iterative FGSM is relatively ineffective. Third, the perturbations we have constructed may not correspond to natural variation in medical images. The study of realistic perturbations that may be more prevalent in MR images (e.g., motion or other types of image artifacts) will be important to study in future work. Other types of contamination that might feasibly arise include random noise in training or testing images, or incorrect labels that are introduced in a random or adversarial manner. Although we hypothesize that the defense strategies proposed in this paper may also be more robust with respect to such perturbations, their efficacy based on this study is unclear. In practice, it may be necessary to devise other defense strategies that are specific to these types of perturbations.

With respect to computational complexity, adversarial training needs to perform forward and backward propagation twice for each batch of data, compared to three forward and two backward processes required for defensive distillation. Therefore, adversarial training is less computationally complex given the same configuration. Moreover, higher values of T in distillation require more iterations for convergence, leading to higher computational costs during the training process. Furthermore, adversarial training is generally more interpretable than defensive distillation: we can check that the perturbed images generated during the training process should indeed be segmented in the same way as the unperturbed images, provided the radius of perturbation is sufficiently small. This provides a natural way to bound the magnitude of $\varepsilon$, whereas it is more difficult to determine the "right" magnitude of T to use without cross-validating the distilled model on test data.

In summary, we have shown that segmentation networks can be adversely affected by the use of targeted attacks which utilize visually minor (and potentially undetectable) modifications to existing images. By adding a small perturbation calculated by FGSM to the input MR image of a patient, normal tissue can be regarded as a tumor by the network. With an increasing interest in applying deep learning techniques to medical imaging data, it is important to understand the ramifications of adversarial inputs (either intentional or unintentional), as these tools may be used in clinical decision-making. We have demonstrated that defensive techniques such as distillation and adversarial training can help combat one-step perturbations added to MR images. As the temperature grows, robustness increases at the cost of computational complexity. Additional work is necessary to continue the exploration and understanding of adversarial approaches that can trick deep learning networks into misclassifying or mislabeling medical images.

**Acknowledgements**

Research reported in this publication was supported by the National Institute of Biomedical Imaging and Bioengineering of the National Institutes of Health under award number R01EB026708

| Attack Type | Dice Coef - tumor core | Dice Coef - enhancing tumor | Dice Coef - whole tumor | Input PSNR | Input SSIM | Input RMSE |
|---|---|---|---|---|---|---|
| No attack | 0.821±0.042 | 0.668±0.253 | 0.748±0.043 | - | - | - |
| FGSM | 0.561±0.077* | 0.370±0.239* | 0.549±0.078* | 27.69±0.28 | 0.646±0.043 | 0.041±0.001 |
| i-FGSM | 0.342±0.087* | 0.230±0.162* | 0.393±0.090* | 27.93±0.13 | 0.470±0.015 | 0.040±0.001 |
| ti-FGSM | 0.461±0.085* | 0.365±0.225* | 0.486±0.082* | 29.48±0.34 | 0.735±0.064 | 0.034±0.001 |

Table 1. Segmentation results for three different attacks: Fast Gradient Sign Method (FGSM), iterative FGSM (i-FGSM), and targeted iterative FGSM (ti-FGSM), quantified via the Dice coefficient of the output segmentation and the PSNR, SSIM, and RMSE of the perturbed input images. For the Dice coefficient measurements, an asterisk (*) indicates statistically significant differences relative to "No attack" at the level $p \leq 0.05$, corrected for multiple comparisons.

| Segmentation Type | Defense Type | Dice Coefficient | | | Dice Coefficient Difference P-value | | |
|---|---|---|---|---|---|---|---|
| | | $\varepsilon = 0$ | $\varepsilon = 0.05$ | $\varepsilon = 0.1$ | $\varepsilon = 0$ | $\varepsilon = 0.05$ | $\varepsilon = 0.1$ |
| Whole Tumor | No Defense | 0.890±0.052 | 0.676±0.189 | 0.609±0.216 | - | - | - |
| | Distillation (T=20) | 0.893±0.053 | 0.741±0.162* | 0.686±0.180* | 0.2287 | 6.628e-4 | 0.0013 |
| | Distillation (T=100) | 0.891±0.057 | 0.807±0.123* | 0.753±0.149* | 0.3714 | 9.011e-9 | 9.970e-8 |
| | Distillation (T=500) | 0.896±0.056* | 0.814±0.153* | 0.759±0.174* | 0.0224 | 1.729e-8 | 3.916e-7 |
| | Distillation (T=5000) | 0.885±0.080 | 0.809±0.161* | 0.758±0.182* | 0.8706 | 4.296e-8 | 1.678e-7 |
| | Adversarial Training-0.01 | 0.887±0.075 | 0.803±0.136* | 0.752±0.158* | 0.5646 | 2.176e-8 | 3.755e-7 |
| | Adversarial Training-0.05 | 0.880±0.061* | 0.803±0.107* | 0.752±0.124* | 0.0457 | 1.991e-7 | 1.007e-6 |
| | Adversarial Training-0.1 | 0.888±0.053 | 0.809±0.128* | 0.765±0.143* | 0.3464 | 6.176e-9 | 7.119e-9 |
| Tumor Core | No Defense | 0.826±0.142 | 0.467±0.242 | 0.374±0.233 | - | - | - |
| | Distillation (T=20) | 0.820±0.174 | 0.603±0.266* | 0.521±0.256* | 0.2137 | 1.982e-6 | 1.950e-5 |
| | Distillation (T=100) | 0.801±0.174* | 0.633±0.259* | 0.543±0.268* | 0.0197 | 2.411e-6 | 9.801e-6 |
| | Distillation (T=500) | 0.823±0.165 | 0.661±0.279* | 0.574±0.289* | 0.9968 | 1.828e-7 | 7.289e-6 |
| | Distillation (T=5000) | 0.810±0.175 | 0.713±0.236* | 0.640±0.247* | 0.5486 | 3.158e-9 | 3.478e-9 |
| | Adversarial Training-0.01 | 0.801±0.181 | 0.639±0.244* | 0.556±0.255* | 0.2137 | 4.699e-8 | 1.334e-6 |
| | Adversarial Training-0.05 | 0.796±0.171* | 0.686±0.230* | 0.629±0.237* | 0.0019 | 4.216e-9 | 7.464e-9 |
| | Adversarial Training-0.1 | 0.793±0.165* | 0.654±0.232* | 0.588±0.241* | 6.628e-4 | 2.478e-9 | 1.250e-8 |
| Enhancing Tumor | No Defense | 0.670±0.295 | 0.380±0.267 | 0.301±0.242 | - | - | - |
| | Distillation (T=20) | 0.672±0.286 | 0.506±0.274* | 0.425±0.261* | 0.0902 | 1.062e-6 | 5.837e-6 |
| | Distillation (T=100) | 0.693±0.238* | 0.517±0.289* | 0.425±0.283* | 0.0016 | 0.0012 | 0.0035 |
| | Distillation (T=500) | 0.704±0.261 | 0.561±0.298* | 0.471±0.294* | 0.3224 | 7.785e-8 | 3.351e-6 |
| | Distillation (T=5000) | 0.660±0.286 | 0.583±0.285* | 0.522±0.275* | 0.1436 | 2.345e-8 | 1.068e-8 |
| | Adversarial Training-0.01 | 0.637±0.289 | 0.525±0.282* | 0.462±0.269* | 0.6259 | 4.602e-8 | 3.648e-6 |
| | Adversarial Training-0.05 | 0.656±0.304* | 0.531±0.283* | 0.445±0.266* | 2.979e-5 | 1.155e-9 | 2.365e-9 |
| | Adversarial Training-0.1 | 0.649±0.283* | 0.518±0.283* | 0.448±0.269* | 0.0057 | 5.064e-9 | 7.594e-8 |

Table 2. Results of different models quantified using the Dice coefficient of label 4 (Enhancing Tumor) when attacked by FGSM with $\varepsilon$ equal to 0, 0.05, and 0.1. An asterisk (*) indicates statistically significant differences at p≤0.05, corrected for multiple comparisons.

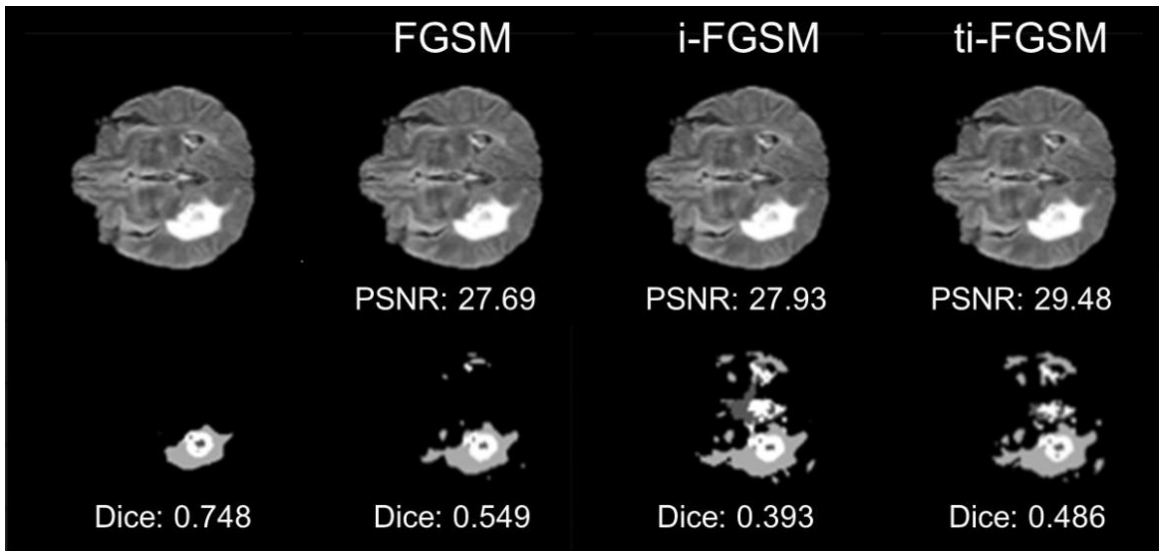

Figure 1. Top row: input images. Bottom row: predicted segmentation for the three adversarial approaches (FGSM, i-FGSM, ti-FGSM) compared to the unperturbed input (far left).

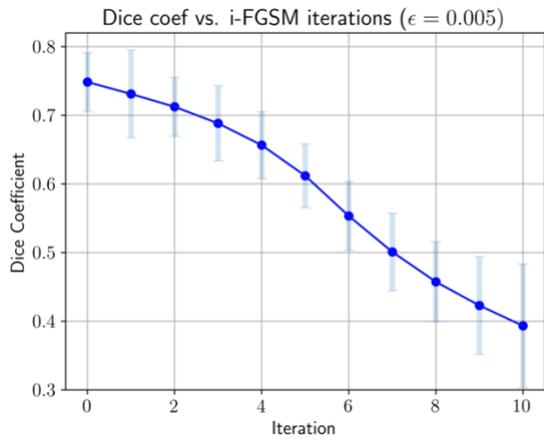 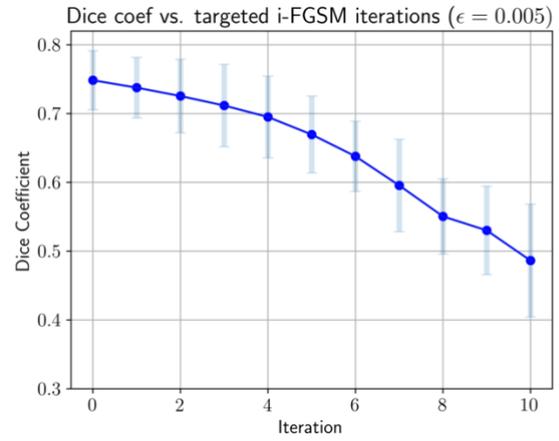

Figure 2. Plots of the Dice coefficient vs. number of Iterations for all study data using i-FGSM and ti-FGSM. Error bars are also shown. As the number of iterations increases, the adversaries become stronger, causing the segmentation output to worsen.

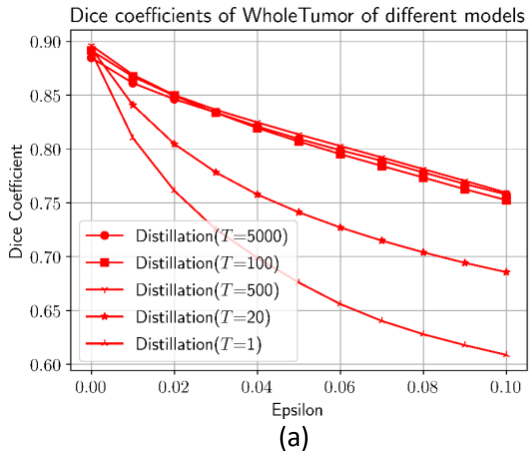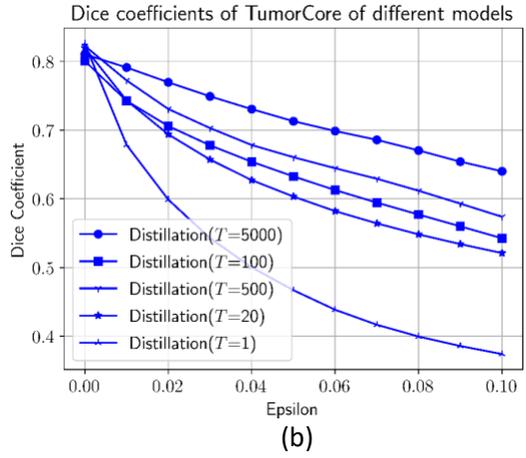

(a) (b)

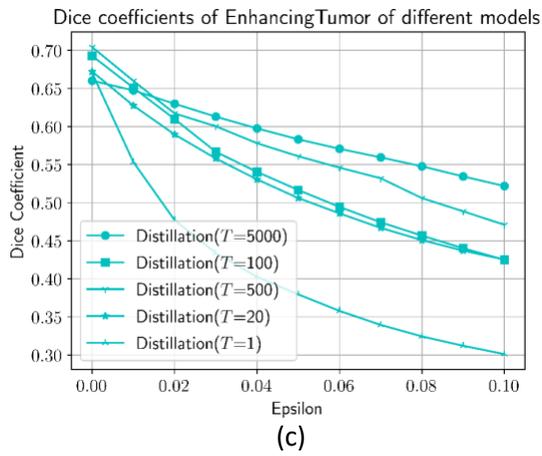

(c)

Figure 3. Performance of distillation. Dice coefficients of (a) "Whole Tumor," (b) "Tumor Core," and (c) "Enhancing Tumor" vs. FGSM with different $e$.

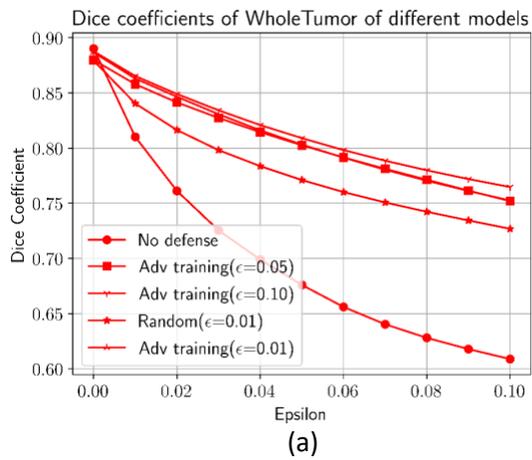
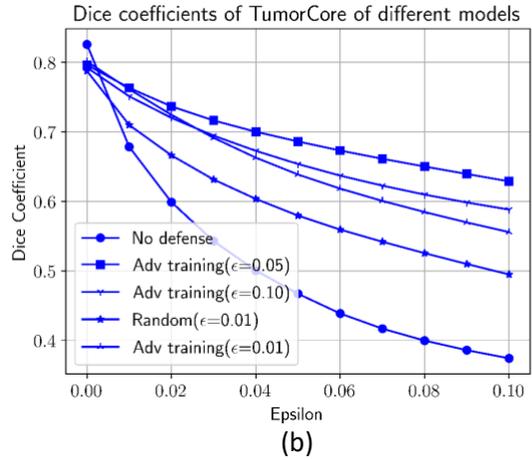

(a) (b)

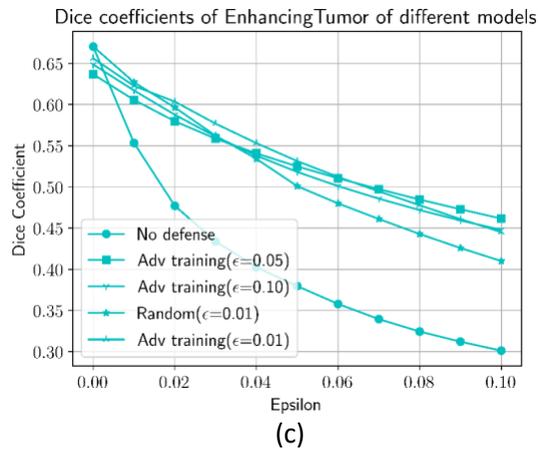

(c)

Figure 4. Performance of adversarial training. Dice coefficients of (a) "Whole Tumor," (b) "Tumor Core," and (c) "Enhancing Tumor" vs. FGSM with different $\epsilon$.

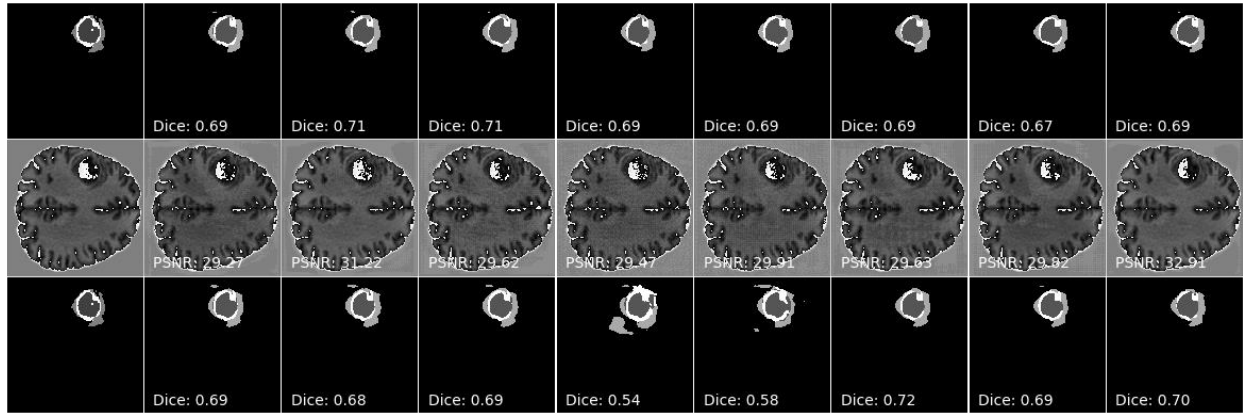

Figure 5. Top row: true labels and predicted segmentations of each model given the original input image. Middle row: original input image and adversarial examples for each model generated by FGSM with $e = 0.03$. Bottom row: true labels and predicted images of each model given their corresponding adversarial examples. Models starting from the 2nd row: adversarial training with $e$=0.05, adversarial training with $e$=0.01, adversarial training with $e$=0.1, model with no defense, distillation with T=20, distillation with T=100, distillation with T=500, and distillation with T=5000.

**Supplemental Materials**

**Description of 3D-Unet**

This model contains 28 convolutional blocks, which includes 3D convolution, instance normalization, and Leaky ReLu layers. To make the network more efficient, it also has residual connections [28]. The network architecture is illustrated in Supplemental Figure 1.

The encoder module contains 15 convolutional blocks with residual connection. In the convolutional layers of these blocks, the size of all kernels is 3 × 3 × 3. The stride is set to 2 if we want the output size of the convolutional layers to be reduced to half of the input size; otherwise, the stride is 1. Since the kernel size is odd, the zero-padding strategy is different. We add 1 column in the left, 2 columns in the right, 1 row at the top, and 2 rows at the bottom. There is a 3D dropout layer with dropout rate of 0.3 between the two orange convolutional blocks to make the training process faster and improve the generalization performance of the model.

In the decoder module, after each upsampling layer, there is one convolution block with kernel size of 3 × 3 × 3 and stride of 1. After each concatenation layer, there are two convolution blocks. The first convolution has same structure as the convolution block after upsampling layers. The kernel size in the second one is 1 × 1 × 1. In the residual links, there are three convolutional layers (colored blue). The kernel size for all these convolutions is 1 × 1 × 1. The number of kernels in each convolution is the same as the number of labels in the ground truth. After the decoder, there is a softmax layer to calculate the probability of each label.

**Details for FGSM**

FGSM perturbs an image X according to the equation

$$X_{adv} = X + \varepsilon \cdot \text{sign}(\nabla_x \text{loss}(X)),$$

where $\varepsilon$ is the upper bound on individual pixel perturbations.

i-FGSM, which is expected to have a higher success rate than FGSM for generating incorrectly classified images, consists of applying FGSM for multiple iterations:

$$X_{k+1} = X_k + \alpha \cdot \text{sign}(\nabla_x \text{loss}(X_k)),$$

where $X_0 = X$.

The ti-FGSM perturbations are defined by the equation:

$$X_{k+1} = X_k + \alpha \cdot \text{sign}(\nabla_x \text{loss}(X_k, \text{target})),$$

where $X_0 = X$, corresponding to iteratively minimizing the loss between the output label and the target label.

**Details for Defensive Distillation**

Let $F(X)$ be the model used for distillation, where $X$ is the input image. The output of $F(X)$ is a 4D array, and for each pixel $(m,n,k)$ of the input image $X$, we have an array of soft labels

$$[F(X)]_{i,m,n,k} = \frac{e^{z_{i,m,n,k}(X)/T}}{\sum_{l=0}^{N-1} e^{z_{l,m,n,k}(X)/T}},$$

where $z_{i,m,n,k}(X)$ is the element with index $(i,m,n,k)$ in the 4D matrix before the activation function is applied, and $N$ is the number of classes in the data set. The temperature $T$ is a constant. If $T = 1$, the above function is the usual softmax.

Let $N_1, N_2, N_3$ be the number of rows, columns, and channels of $X$. Given the one-hot truth matrix $Y$ with dimensions $(N, N_1, N_2, N_3)$, the Dice coefficient is calculated as

$$D(X,Y) = \frac{1}{N} \sum_{i=1}^{N-1} \frac{2 \sum_{m=0}^{N_1-1} \sum_{n=0}^{N_2-1} \sum_{k=0}^{N_3-1} [F(X)]_{i,m,n,k} \cdot [Y]_{i,m,n,k} + \gamma}{\sum_{m=0}^{N_1-1} \sum_{n=0}^{N_2-1} \sum_{k=0}^{N_3-1} [F(X)]_{i,m,n,k} + \sum_{m=0}^{N_1-1} \sum_{n=0}^{N_2-1} \sum_{k=0}^{N_3-1} [Y]_{i,m,n,k} + \gamma},$$

where $\gamma$ is a small positive real number. Note that by default, we set the normal pixels (i.e., the background of the image and the part of tissue that does not have any disease) to class 0 ($i=0$ here). Then the loss function is defined according to the Dice coefficient

$$\text{loss}(X,Y) = 1 - D(X,Y).$$

Let $F^d$ be the distilled network, which has the same architecture as $F$. When training $F^d$, the only difference is that instead of the one-hot matrix $Y$, we use the output $F(X)$ from the first trained network $F$. We define

$$D(X, F(X)) = \frac{1}{N} \sum_{i=1}^{N-1} \frac{2 \sum_{m=0}^{N_1-1} \sum_{n=0}^{N_2-1} \sum_{k=0}^{N_3-1} [F^d(X)]_{i,m,n,k} \cdot [F(X)]_{i,m,n,k} + \gamma}{\sum_{m=0}^{N_1-1} \sum_{n=0}^{N_2-1} \sum_{k=0}^{N_3-1} [F^d(X)]_{i,m,n,k} + \sum_{m=0}^{N_1-1} \sum_{n=0}^{N_2-1} \sum_{k=0}^{N_3-1} [F(X)]_{i,m,n,k} + \gamma},$$

and define the loss function of the distilled network as $\text{loss}(X, F(X)) = 1 - D(X, F(X))$.

For adversarial training [8], the goal is to determine a model that minimizes the population risk:

$$\min_{\theta} E_{(x,y) \sim D} \left[ \max_{\delta \in S} L(\theta, x+\delta, y) \right],$$

where $S$ is the set of allowed perturbations, $D$ is the data distribution, and $L$ is the loss function. In practice, the set $S$ is often defined to be the $l_\infty$-ball with radius $\varepsilon$, meaning that each pixel can be perturbed by at most $\varepsilon$. To minimize the expectation above, a natural strategy is to perform gradient descent on the adversarial loss function. It may be shown that the gradient of the adversarial loss function at $X$ is identical to the gradient of the usual loss function evaluated at the "worst case" point in the neighborhood of $X$. Identifying this worst-case point is not computationally feasible, so a popular alternative is to use an adversarial attack (e.g., FGSM), and then train the model by evaluating the gradient at the adversarial example. In practice, we often use the adversarial objective function based on FGSM as an effective regularizer [7]:

$$\tilde{L}(\theta, x, y) = \alpha L(\theta, x, y) + (1-\alpha) L(\theta, x + \varepsilon \text{sign}(\nabla_x L(\theta, x, y)), y),$$

where $\alpha$ is the weight factor. When $\alpha = 0.5$, it works well, though there may be other values that can achieve better performance. Therefore, the goal of adversarial training is to minimize $\tilde{L}(\theta, x, y)$ over the training dataset.

Due to memory limitations, we implement the minimization in an iterative way. For each batch of training data, we first generate their adversarial examples based on the current model, and do forward and backward propagation by using these adversarial examples. Then the model is updated according to the original batch of data.

**Training Details**

For data augmentation, we apply uniform perturbations of radius of 0.01 in infinity norm to the input data. During training, for each batch of input data, we first train the model with the perturbed images, and then train the model with the clean images, in order to obtain a fair comparison to adversarial training.

For data preprocessing, we apply N4 bias field correction [29] and global standardization. The 3D images were resized to 128 x 128 x 128 to match the input shape of the 3D UNet. For training , we  used a batch size of 1, and the number of epochs to be 100. The Adam optimizer was used with a learning rate of 1e-4. When training the distilled model with a large temperature, more iterations were required, so we increased the number of epochs to be 400 and the learning rate to be 5e-4. To allow improved generalization, data augmentation was performed on the 3D images, which includes rotation within the axial slices, flips, and matrix transposes.

**Effect of i-FGSM on Image Input Quality**

In Supplemental Figure 2 we show plots of the average PSNR vs. number of iterations in i-FGSM and ti-FGSM. The decrease in PSNR is expected; however, note that the average PSNR is still reasonably large, implying that the quality of the perturbed images is relatively high. Additionally, as seen in Fig. 1, the effects are barely discernible, suggesting that PSNR (and the other image quality metrics of SSIM and RMSE) are sensitive to the FGSM attacks.

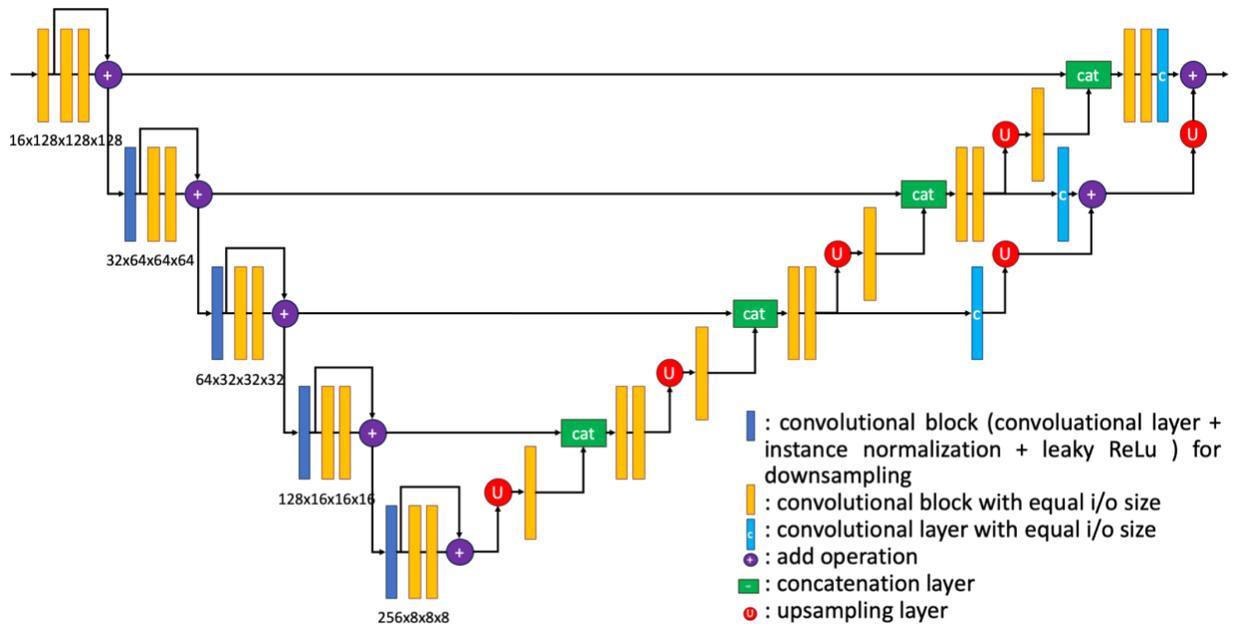

Supplemental Figure 1. The 3D U-Net architecture.

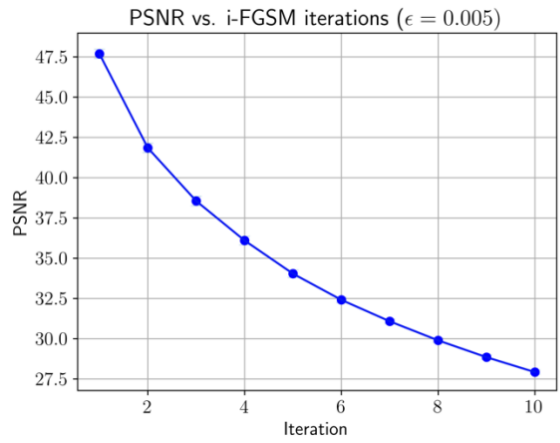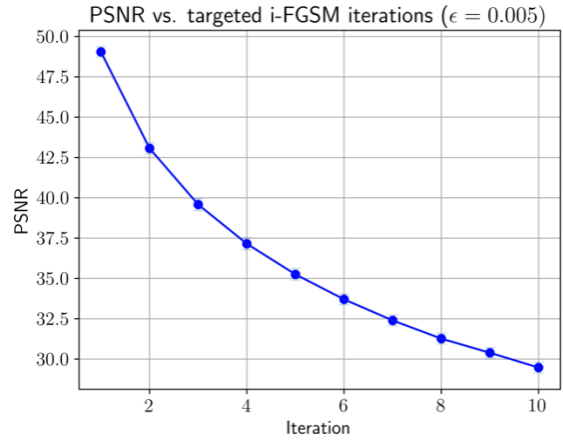

Supplemental Figure 2. Average PSNR vs. number of iterations in i-FGSM and ti-FGSM.